\documentclass[conference]{IEEEtran}
\IEEEoverridecommandlockouts
\usepackage{cite}
\usepackage{amsmath,amssymb,amsfonts}
\usepackage{algorithmic}
\usepackage{graphicx}
\usepackage{textcomp}
\usepackage{xcolor}
\usepackage{stfloats}
\usepackage{tikz}
\def\BibTeX{{\rm B\kern-.05em{\sc i\kern-.025em b}\kern-.08em
    T\kern-.1667em\lower.7ex\hbox{E}\kern-.125emX}}

\newcommand\copyrighttext{%
  \footnotesize \textcopyright 2025 IEEE.  Personal use of this material is permitted.  Permission from IEEE must be obtained for all other uses, in any current or future media, including reprinting/republishing this material for advertising or promotional purposes, creating new collective works, for resale or redistribution to servers or lists, or reuse of any copyrighted component of this work in other works.}
\newcommand\copyrightnotice{%
\begin{tikzpicture}[remember picture,overlay]
\node[anchor=south,yshift=10pt] at (current page.south) {\fbox{\parbox{\dimexpr\textwidth-\fboxsep-\fboxrule\relax}{\copyrighttext}}};
\end{tikzpicture}%
}

\begin{document}

\title{Nonlinear Co-simulation for Designing Kinetic Inductance Parametric Amplifiers}

\author{
\IEEEauthorblockN{
Likai Yang\IEEEauthorrefmark{1},
Yufeng Wu\IEEEauthorrefmark{2},
Chaofan Wang\IEEEauthorrefmark{2},
Mingrui Xu\IEEEauthorrefmark{2},
Hong X. Tang\IEEEauthorrefmark{2},
Mohamed A. Hassan\IEEEauthorrefmark{1},
Eric T. Holland\IEEEauthorrefmark{1}
}
\IEEEauthorblockA{
\IEEEauthorrefmark{1}Keysight Technologies, USA \\
\IEEEauthorrefmark{2}Department of Electrical Engineering, Yale University, New Haven, USA \\
Email: likai.yang@keysight.com, hong.tang@yale.edu \\
}
}

\maketitle
\copyrightnotice
\begin{abstract}
Kinetic inductance parametric amplifiers (KIPAs) have been widely studied for small-signal detection in superconducting quantum circuits. In this work, we demonstrate the modeling of a niobium nitride nanowire based KIPA using electromagnetic (EM) and circuit co-simulation, and compare the outcomes with experimental results. EM analysis is first performed on the device layout, taking into account the linear part of the kinetic inductance. The results are then integrated into a harmonic balance circuit simulator, in which the current-dependent inductance is modeled by representing the nanowire as a nonlinear inductor. Both linear and nonlinear responses of the device, including temperature-dependent resonance spectra and parametric gain, are extracted and show good agreement with experiments. We further show that when the KIPA operates as a degenerate amplifier, its phase-sensitive behavior can be accurately reproduced by the simulation. Our technique can serve as a valuable enabler for the simulation and design of quantum parametric amplifiers and superconducting kinetic inductance devices.
\end{abstract}


\section{Introduction}
Superconducting quantum technologies routinely deal with weak microwave signals whose power is orders of magnitude below the thermal noise at room temperature, to enable applications such as qubit readout \cite{wallraff2004strong}, spin detection \cite{wang2023single}, and quantum sensing \cite{degen2017quantum}. This necessity thus drives the development of cryogenic parametric amplifiers \cite{ho2012wideband,macklin2015near}, which exploit the Kerr nonlinearity for realizing near-quantum-limited amplification. The most common type of superconducting parametric amplifiers utilizes the nonlinearity of aluminum-based Josephson Junctions (JJs). These devices, however, face limitation in operating at higher frequencies and elevated temperatures due to the relatively low superconducting energy gap and critical temperature of aluminum. Their lack of resilience to external magnetic field also prevents direct integration with systems such as solids state spins \cite{wang2022high} and topological qubits \cite{aasen2025roadmap}. Alternatively, these obstacles can be lifted by using kinetic inductance parametric amplifiers (KIPA), in which the current-dependent inductance of superconductors is leveraged as the source of nonlinearity. This is because many high kinetic inductance materials, such as niobium nitride (NbN) and niobium titanium nitride (NbTiN), possess much higher critical temperatures over 10\,K and critical field of multiple Tesla \cite{baskaran2014high,makise2010characterization}. To date, KIPA devices capable of operating at millimeter-wave frequencies \cite{anferov2020millimeter} and liquid-helium temperature \cite{malnou2022performance,frasca2024three} have been demonstrated. Their functionality under magnetic field close to one Tesla has been established and utilized for sensitive spin detection \cite{xu2023magnetic,vine2023situ}. Broadband traveling-wave KIPAs \cite{malnou2021three,faramarzi20244} have also been realized and shown comparable performance with their counterparts based on Josephson Junctions. Furthermore, it is also proposed that the nonlinearity of kinetic inductance can be leveraged to build high-frequency qubits \cite{faramarzi2021initial}. In light of these advantages, kinetic inductance based devices would serve as valuable resources in the application of superconducting quantum circuits.

KIPAs usually consist of one or more highly-nonlinear elements, such as narrow superconducting wires \cite{joshi2022strong}, in conjunction with linear components that form transmission lines or resonators. It is thus crucial to capture both the linear and nonlinear components of kinetic inductance in the simulation and design of KIPAs. Its linear part acts as additional inductance on top of the geometric inductance, thereby modifying the cavity resonance frequency or the impedance of the transmission line. This effect should be taken into account in the electromagnetic (EM) analysis, which is widely adopted for modeling the response of linear devices. On the other hand, extracting nonlinear properties of parametric amplifiers can be realized with circuit simulation \cite{guarcello2022modeling,peng2022x,sweetnam2022simulating,shiri2024modeling}, which is commonly implemented by constructing a lumped-element circuit model of the device. This approach, however, may fall short in properly capturing distributed kinetic inductance and reproducing complicated device structures. Given these considerations, it is highly desirable to combine linear EM simulation with nonlinear circuit simulation, a technique often referred to as co-simulation, for accurate and efficient modeling of KIPAs.

In this work, we implement the co-simulation method to model a cavity-based KIPA made from thin-film NbN, whose nonlinearity arises from a 15\,nm-wide nanowire. The simulations are conducted with Keysight Advanced Design System (ADS), and the outcomes are compared with experimental characterization results. We begin with an EM analysis of the device layout. The linear kinetic inductance is incorporated by defining a London penetration depth in the superconducting material model in ADS. The extracted external coupling rate and temperature-dependent resonance frequency of the cavity show good agreement with experiments. We then perform nonlinear co-simulation by replacing the nanowire with a nonlinear inductor circuit element and integrating it with EM-simulated S-parameter in a harmonic balance simulator. Key nonlinear behaviors of the device including resonance bifurcation and parametric gain can be successfully reproduced and match well with measured results. Thanks to the capability to do comprehensive sweep of parameters in the simulation, we perform a two-dimensional sweep of pump power and frequency to show that the KIPA can achieve high gain in a 180\,MHz frequency range. Furthermore, the simulation accurately predicts the phase-sensitive behavior of the KIPA when operating as a degenerate amplifier \cite{parker2021near}, which is of particular interest for high-sensitivity detection and squeezed state generation. We note that detailed experimental studies of this device structure and part of the measurement data shown here have been published in our previous works \cite{xu2023magnetic,xu2024radiatively}.

\section{Device Basics}

\begin{figure}[t]
\centering
\includegraphics[width=0.5\textwidth]{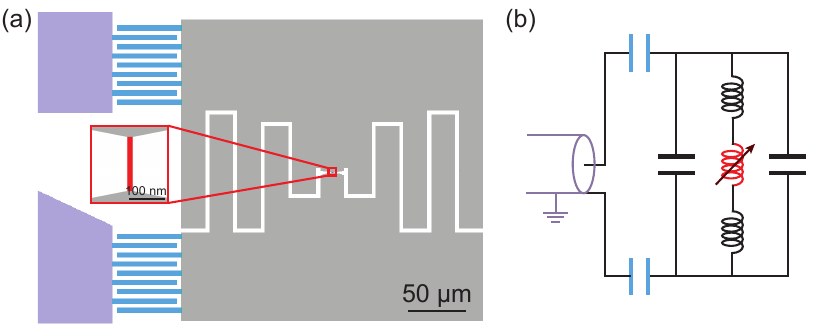}
\caption{Device schematics. (a) The layout of the KIPA device, featuring a LC resonator coupled to a coplanar waveguide via a pair of interdigital capacitor. A zoom-in view of the nanowire is also shown. Each component is color coded to correspond with the circuit diagram. (b) A simplified circuit diagram representing the device. The nanowire can be viewed as a nonlinear inductor.}
\label{fig1}
\end{figure}

The NbN thin film used for our KIPA devices are grown by atomic layer deposition (ALD) on a high-resistivity silicon substrate. To obtain high kinetic inductance, the film thickness is set to be 4\,nm. Its sheet inductance and critical temperature is measured in a separate characterization to be 180\,pH/$\blacksquare$ and 9\,K, respectively. The layout of the device is shown in Fig.~\ref{fig1}(a). A LC resonator is formed by connecting two NbN pads with a nanowire, whose width and length are determined to be 15\,nm and 150\,nm, respectively. It is coupled to a 50\,$\Omega$ coplanar waveguide by a pair of interdigitated capacitors and the cavity resonance can be accessed by measuring the reflection. To fabricate the device, the NbN thin film is first patterned with electron-beam lithography and subsequently etched by a reactive ion etching system. The characterization of the device is done by cooling it down to 10\,mK in a dilution refrigerator. The details about device fabrication and experimental setup can be found in Ref.~\cite{xu2023magnetic}.

The device structure can be understood by a simplified circuit schematics, as sketched in Fig.~\ref{fig1}(b). The inductance of the resonator is constructed by a nonlinear inductor, representing the nanowire, together with the linear inductors from the geometric and kinetic inductance of the two pads. They are connected in parallel with the parallel capacitors formed by the slits. The current-dependent kinetic inductance of the nanowire can be described as \cite{semenov2020effect}
\begin{equation}
L_k=L_{k0}\left[1+(\frac{I}{I_*})^2\right],
\label{eq1}
\end{equation}
where $L_{k0}$ is the linear kinetic inductance and $I_*$ is the characteristic current that determines the scaling of the nonlinearity. The quadratic dependence of kinetic inductance on the current gives rise to the Kerr nonlinearity, thus facilitates parametric amplification via a four-wave mixing process. The value of the characteristic current would depend on the material properties and the geometry of the wire cross section, with the formula \cite{shu2021nonlinearity}
\begin{equation}
I_*=wt\kappa_*\sqrt{\frac{N_0\Delta^2}{\mu_0\lambda_L^2}}.
\label{eq2}
\end{equation}
Here, $N_0$, $\Delta$, and $\lambda_L$ are the density state at the Fermi level, superconducting energy gap, and London penetration depth, respectively. The $w$ and $t$ represent the width and thickness of the wire with a rectangular cross section, while $\kappa_*$ is a scaling constant of order 1. The fact that $I_*$ is inversely proportional to the conductor width justifies our estimation that all the nonlinearity comes from the nanowire, while the kinetic inductance from the rest of the device can be viewed as constant. 

\section{Linear EM Analysis}

\begin{figure}[t!]
\centering
\includegraphics[width=0.5\textwidth]{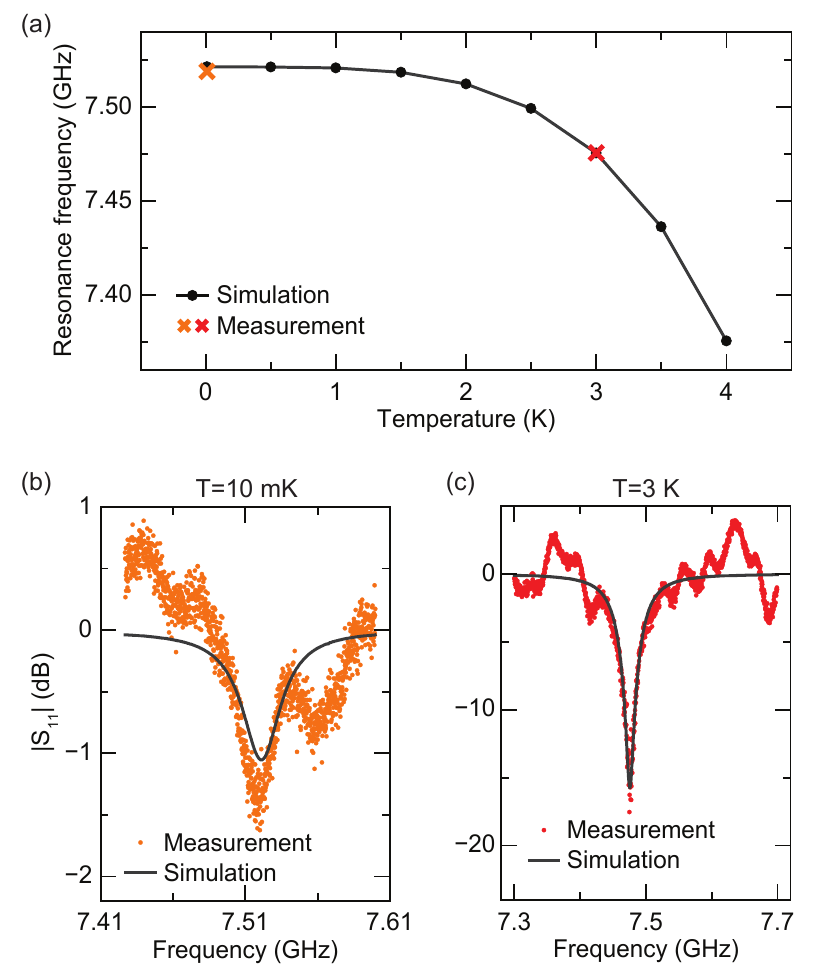}
\caption{EM analysis results. (a) Simulated resonance frequency as a function of temperature. The dots represent simulated temperatures and they are connected to guide visualization. The orange and red cross marks correspond to the measured resonance frequency at 10\,mK and 3\,K, respectively. (b) (c) Simulated and measured resonance spectra at the two different temperatures. The resonance linewidth and extinction match well between the two. This suggests that the simulation accurately captures the external coupling rate. The intrinsic loss is reproduced by artificially set the material loss to match with experiments.}
\label{fig2}
\end{figure}

We first import the device layout into ADS and use a method of moments (MoM) solver to perform an EM analysis. We take advantage of the embedded superconductor model in ADS, in which the linear kinetic inductance is taken into account by defining the London penetration depth of the material. To be specific, in ADS the sheet kinetic inductance is calculated as \cite{brorson1994kinetic,kerr1999surface}
\begin{equation}
L_{k\blacksquare}=\mu_0\lambda_L(T)\mathrm{coth}(\frac{t}{\lambda_L(T)}).
\label{eq3}
\end{equation}
Here, $\lambda_L(T)$ is the temperature-dependent penetration depth. It can be implied from the 0\,K value $\lambda_L(0)$, using
\begin{equation}
\lambda_L(T)=\lambda_L(0)/\sqrt{1-(\frac{T}{T_c})^4}.
\label{eq4}
\end{equation}
In our simulation, we set $\lambda_L(0)=731$\,nm, film thickness $t=4$\,nm, and critical temperature $T_c=9$\,K based on experimental characterization results. By varying the simulation temperature, the temperature-dependent resonance frequency can be extracted, as shown in Fig.~\ref{fig2}(a). Due to experimental limitations, the device was only measured at two discrete temperatures, i.e. 10\,mK in the dilution refrigerator and 3\,K in a close-loop helium cryostat. The results are shown by the cross marks in Fig.~\ref{fig2}(a). The simulated (measured) resonance frequency is 7.5214\,GHz (7.5191\,GHz) at 10\,mK and 7.4756\,GHz (7.4757\,GHz) at 3\,K, respectively. The deviation of the simulated resonance frequencies from and experimental results remains well below 5\,MHz. We note that the temperature shift of frequency for our device is relatively small within 100\,MHz, due to the high $T_c$ of NbN. However, this effect could be more prominent for high kinetic inductance materials with lower $T_c$ such as titanium nitride and oxidized aluminum.

The comparison between simulated and measured resonance spectra at 10\,mK and 3\,K are illustrated in Fig.~\ref{fig2}(b) and (c), respectively. At 3\,K, the resonance feature has a high extinction over 15\,dB, suggesting that the device is nearly critical-coupled. When cooling down to 10\,mK, the intrinsic quality factor (Q) increases while the external coupling rate remains constant. As a result, the resonance becomes over-coupled with an external coupling rate of $\kappa_e=2\pi\times57$\,MHz and an intrinsic loss rate $\kappa_i=2\pi\times1.8$\,MHz. On the other hand, the external coupling rate extracted from the simulation is $2\pi\times55$\,MHz, which agrees with experiments. In designing resonator-based parametric amplifiers, it is favorable to work in the over-coupled regime for high power retrieval efficiency. Thus, the ability to accurately determine the coupling rate with EM analysis further highlights its importance in designing KIPA devices. The material loss in the simulation is manually tuned at different temperature to match the measured intrinsic Q.

\section{Nonlinear EM and Circuit Co-simulation}

\begin{figure*}[t]
\centering
\includegraphics[width=0.98\textwidth]{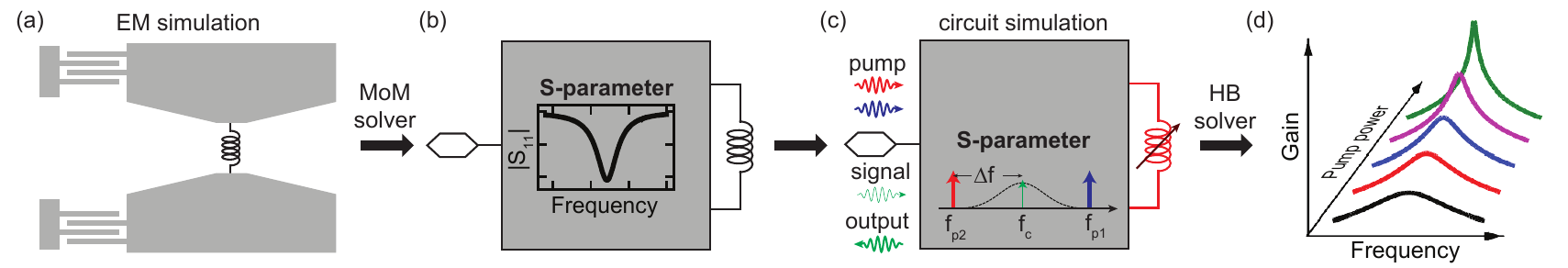}
\caption{Setting up EM and circuit co-simulation. (a) EM simulation is first performed by replacing the nanowire with a component model of a linear inductor in the device layout (drawing not to scale). (b) The S-parameter is extracted with a method of moments (MoM) solver. (c) The linear response is then incorporated into circuit simulation where the nanowire is represented by a nonlinear inductor instead. A two-tone pump scheme is used for generating parametric gain. (d) Nonlinear simulation is performed with a harmonic balance (HB) solver. Parameters including signal frequency, pump power, and pump frequency, etc. are swept to extract the KIPA response.}
\label{fig3}
\end{figure*}

To model the nonlinear properties of the KIPA, we implement a co-simulation method combining EM and circuit simulators. The process to set up the nonlinear co-simulation is sketched in Fig.~\ref{fig3}. As described above, we take the assumption that the device nonlinearity is dictated only by the nanowire, while the kinetic inductance from the rest of the device is linear. With this approximation, we replace the nanowire with a linear inductor circuit component and rerun the EM simulation. The value of the linear inductor is determined to be $L_{k0}=1.705$\,nH in a separate EM analysis of the nanowire. By doing so, a multi-port S-parameter can be extracted from EM simulation and fed into the circuit simulator to represent the linear portion of the device. Meanwhile, the linear inductor representing the nanowire is substituted by a current-dependent nonlinear inductor following Eq.~\ref{eq1}. The characteristic current is set to be $I_*=3.985$\,{\textmu}A. We note that while $I_*$ cannot be directly inferred from EM simulation, once it is experimentally characterized for a single instance of nanowire geometry, it can be straightforwardly generalized using Eq.~\ref{eq2} for nanowires of different cross sections. It is thus a universal parameter that can be used in designing different device configurations on the same material platform. It can be seen as an analogy of the critical current in JJ based devices, and is usually characterized for a single device and applied to the subsequent design flow.

After setting up the device model, we study the nonlinear behaviors of the device using a harmonic balance simulator. The harmonic balance method has been widely adopted for large signal analysis of highly nonlinear microwave circuits. Its accuracy and reliability have been proven by showing good agreement with analytic results from solving coupled mode equations \cite{sweetnam2022simulating}. To facilitate parametric amplification, we implement a two-tone pump scheme as shown in Fig.~\ref{fig3}(c). Two pump tones at $f_{p1}=f_c+\Delta f$ and $f_{p2}=f_c-\Delta f$ with same power are applied, where $f_c$ is the center frequency and $\Delta f$ is the detuning. A small signal with frequency $f_s$ is added and swept around $f_c$ to extract the gain spectrum. The harmonic balance method finds the device output at different mixing terms of $f_{p1}$, $f_{p2}$, and $f_s$ as well as their harmonics. e.g. The mixing terms for signal and idler would be at $f_s$ and $f_{p1}+f_{p2}-f_s$, respectively. The gain is then calculated as the ratio between the output power of the mixing term at $f_s$ and the input signal power. Also, since the pump power of the two pump tones are identical, the pump power described hereafter will be referring to the power of one of the pump tones. In the following, we discuss different results from the nonlinear co-simulation and compare them with experimental data. We note that the nonlinear properties shown below is measured six months after the linear characterization in Fig.~\ref{fig2}. During this time interval, thin oxidization layer was formed on NbN surface, effectively reduced the film thickness by a small amount. This resulted in a slight increase in sheet inductance thus shifted the 10\,mK resonance frequency from 7.52\,GHz to 7.45\,GHz. This effect is incorporated into the simulation by reducing the film thickness from $t=4$\,nm to $t=3.96$\,nm, which is used in all of the following analysis. 

\begin{figure}[b!]
\centering
\includegraphics[width=0.5\textwidth]{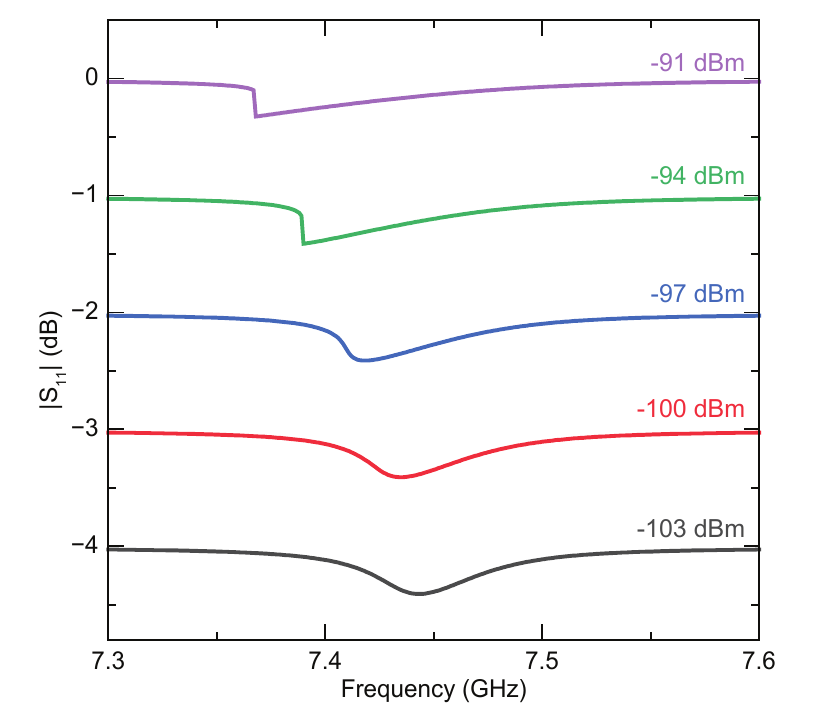}
\caption{KIPA response under different signal power. The simulated $S_{11}$ of the device under varying signal power (no pump applied) is shown and artificially offset by 1\,dB for visibility. The resonance undergoes bifurcation when the signal power is increased over -97\,dBm.}
\label{fig4}
\end{figure}

We first simulate the device response under different signal power, without introducing the pumps. The distortion of resonance shape under high power drive is a typical behavior of nonlinear oscillators exhibiting Kerr nonlinearity. Over a certain threshold power, the resonator will go through bifurcation and exhibits bistability. In the frequency response, this effect can be seen by an abrupt change in reflection amplitude at a certain frequency, depending on the drive power. The simulated results are illustrated in Fig.~\ref{fig4}, where device responses at different drive powers are shown and artificially offset by 1\,dB for better visibility. We can clearly see that the resonance exhibits a simple Lorentzian shape for low drive power of -103\,dBm. However, when the drive power is increased over -97\,dBm, which is determined to be the bifurcation threshold power, the resonance is shifted to lower frequency and has a triangle shape. The bifurcation threshold power is also widely used in experimental study to characterization the strength of Kerr nonlinearity \cite{eichler2014controlling}. Here, the fact that the bifurcation behavior is reproduced in the simulation serves as a validation for our modeling approach.

\begin{figure*}[t]
\centering
\includegraphics[width=0.98\textwidth]{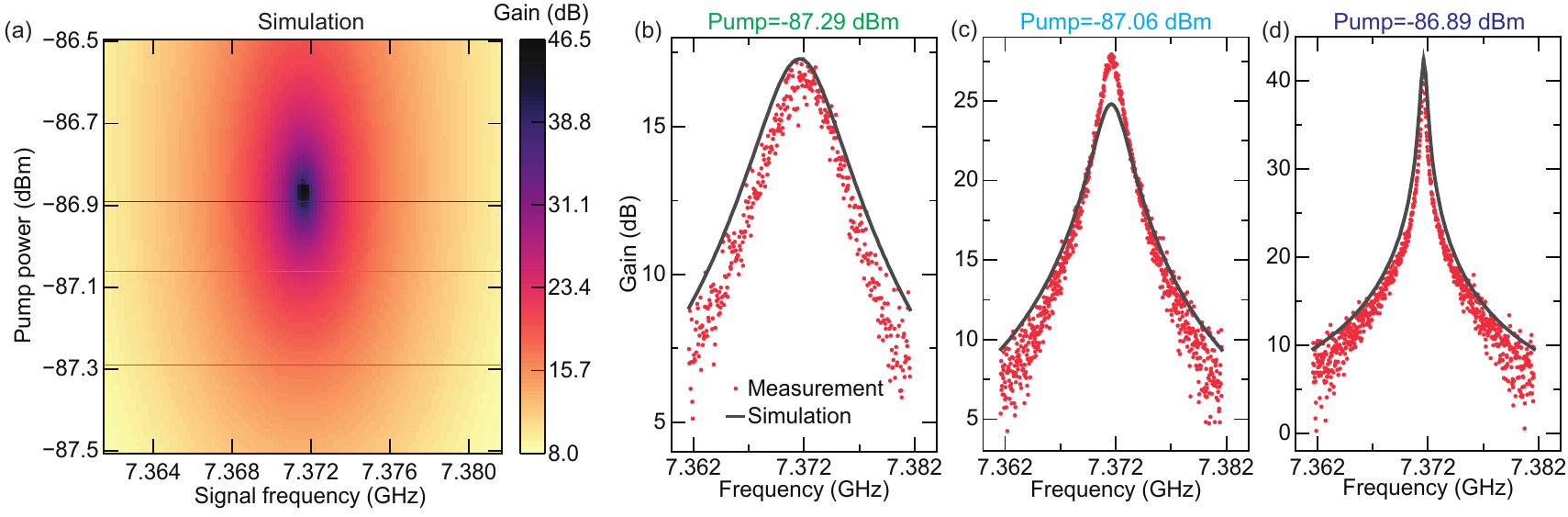}
\caption{Simulated gain spectra at varying pump power and comparison with experiments. The frequencies of two pumps are fixed at 7.3716\,GHz$\pm$133.5\,MHz. (a) Two-dimensional sweep of parametric gain by varying the pump power and signal frequency in the simulation. The green, blue, and purple lines correspond to the pump powers used in the experiments. (b),(c),(d) Comparison between simulated and measured gain spectra at three different pump power. As the pump power is increased from -87.29\,dBm to -86.89\,dBm, the peak gain is increased while the 3-dB gain bandwidth is reduced.}
\label{fig5}
\end{figure*}

We then proceed to simulate the parametric gain under the two-tone pump scheme described above. To do so, we fix the center frequency at $f_c=7.3716$\,GHz and the detuning at $\Delta f=133.5$\,MHz. The pump power is then varied and the small-signal gain at signal frequencies around $f_c$ is extracted. We note that for the simulations and experiments discussed here, we operate the KIPA in the non-degenerate mode, i.e. the signal and idler are not at the exact same frequency. The extracted gain is thus phase-preserving. The results from this two-dimensional sweep are shown in Fig.~\ref{fig5}(a). We can clearly see that the gain spectrum exhibits a symmetric shape and peaks at signal frequency $f_c$. A maximum gain of 46.5\,dB is achieved with a pump power of -86.86\,dBm and would reduce if the pump power is further increased. When the pump power is moderately decreased to -87.5\,dBm, a dramatic reduction in peak gain to 8.0\,dB will be seen, showcasing the importance of finding the optimal working point for the KIPA. Experimentally, the pump power at the device was calibrated using a variable temperature stage \cite{xu2023magnetic} and the gain spectra were measured at three discrete power levels, as indicated by the colored horizontal lines in Fig.~\ref{fig5}(a). The simulated and measured gain under the same pump power are compared in Fig.~\ref{fig5}(b)-(d). With -87.29\,dBm pump power, the simulated (measured) peak gain is 17.3\,dB (16.5\,dB) and the 3-dB bandwidth is 8.0\,MHz (7.5\,MHz). At intermediate pump power of -87.06\,dB, the simulated (measured) peak gain and 3-dB bandwidth are 24.8\,dB (27.5\,dB) and 3.3\,MHz (2.5\,MHz), respectively. When the pump power is further increased to -86.89\,dBm, the simulated (measured) peak gain is increased to 42.0\,dB (42.4\,dB) while the 3-dB bandwidth is reduced to 0.5\,MHz (0.3\,MHz). There thus exists a trade-off between the maximally achievable gain and its bandwidth. The discrepancy of gain-bandwidth product between simulations and experiments is less than 41\,\% for all cases. The mismatch can potentially be attributed to the inaccuracy in modeling the resonance linewidth, which impacts the gain bandwidth, as well as the error in measuring the exact on-chip pump power.

\begin{figure}[b!]
\centering
\includegraphics[width=0.5\textwidth]{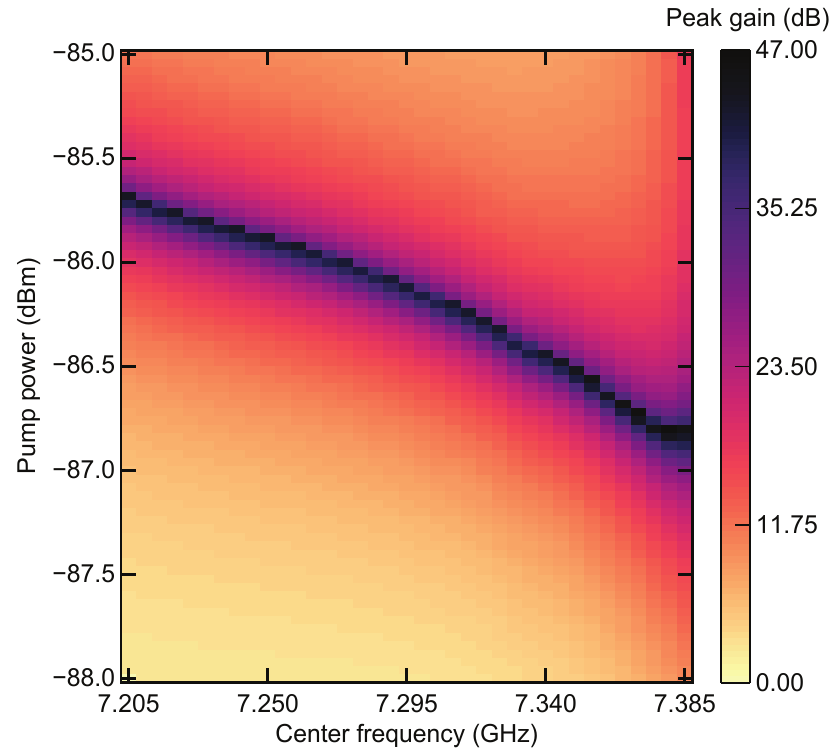}
\caption{Simulated peak gain of the KIPA when sweeping the pump power and center frequency $f_c$. The detuning is kept constant at $\Delta f=133.5$\,MHz. Gain over 45\,dB can be achieved for a frequency span of 180\,MHz, which is around three times the resonator linewidth. The pump power required to achieve maximum gain increase linearly with lower center frequency.}
\label{fig6}
\end{figure}

Due to the correlation between gain and bandwidth as well as their delicate dependency on pump power and frequency, the nonlinear simulation can serve as a valuable tool for comprehensive analysis of device working conditions and provide guidance on experimental set points. This is because multi-dimensional sweeping and optimization of input parameters can be easily realized in our simulation. Here, we leverage this advantage to find the peak gain under varying pump power and frequency, thus extracting the usable frequency range of the KIPA. While keeping the detuning $\Delta f$ constant at 133.5\,MHz, the pump center frequency $f_c$ is swept from 7.205\,GHz to 7.385\,GHz and pump power is swept from -88.0\,dBm to -85.0\,dBm. The simulated peak gain at center frequency is plotted in Fig.~\ref{fig6}. As shown by the results, the KIPA can achieve peak gain over 45\,dB within this frequency range. This corresponds to an operable bandwidth of 180\,MHz for our KIPA devices, which is about three times its resonance linewidth. The required pump power to achieve the highest gain, on the other hand, increases linearly with detuning as the center frequency is shifted further away from the resonance frequency. 

The KIPA can also work as a degenerate amplifier for phase-sensitive amplification, which is particularly useful for increasing measurement sensitivity with noiseless amplification and squeeze state generation. A phase-preserving amplifier amplifies the two quadratures of the signal equally, meanwhile adding at least 1/4 noise photon to each quadrature, known as the quantum limit. The phase-sensitive mode of operation, however, can amplify one quadrature while deamplify the other, allowing the amplification of a single quadrature without added noise. Similar process could be applied to vacuum fluctuation to deamplify or squeeze the fluctuation on one quadrature, thus improve the noise performance in small-signal detection. In our scheme, this happens when the signal frequency is exactly at $f_c$ so that the signal and idler become degenerate. The interference between them thus facilitates phase-sensitive amplification, indicated by a change in gain amplitude with respect to the phase of the signal and the pumps. i.e. The phase-sensitive gain is dictated by the phase difference between signal and idler, denoted by
\begin{equation}
\Delta \varphi=2\varphi_s-(\varphi_{p1}+\varphi_{p2}),
\label{eq5}
\end{equation}
where $\varphi_s$ is the phase of the signal and $\varphi_{p1},\varphi_{p2}$ are the phase of the two pumps, respectively. The transition from maximum amplification to maximum deamplification of the signal will thus take place for every $\pi$ phase shift of $\Delta \varphi$. 

\begin{figure}[t]
\centering
\includegraphics[width=0.5\textwidth]{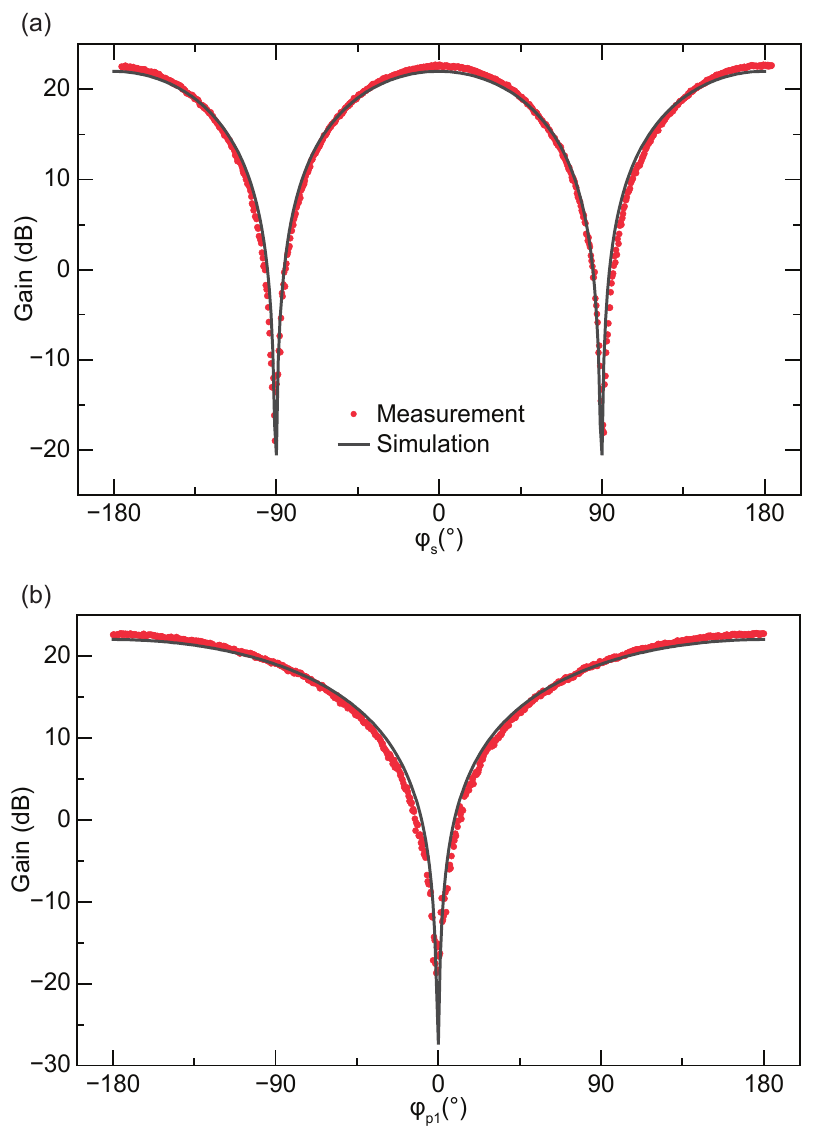}
\caption{Phase-sensitive amplification when the KIPA operates in the degenerate mode, i.e. signal frequency is at $f_c$. The pump parameters used in the simulations are $f_c=7.3617$\,GHz, $\Delta f=133.5$\,MHz, and pump power -87.35\,dBm. The gain amplitude will change with respect to the phase of the signal and the pumps, due to the interference between signal and idler. Both simulated and measured gain are shown and compared. (a) Varying the signal phase $\varphi_s$. (b) Varying the phase of one of the pumps $\varphi_{p1}$.}
\label{fig7}
\end{figure}

To reproduce this behavior, we fix the signal frequency at $f_c=7.3716$\,GHz and run the harmonic balance simulation. Furthermore, instead of taking only the mixing term of the signal to be the output, we take the sum of the signal ($f_c$) and idler ($f_{p1}+f_{p2}-f_c$) mixing terms as they are degenerate. The phase of the signal or the pumps can then be swept to see the variation in gain. We fix the pump power to be -87.35\,dBm and first plot the simulated as well as measured gain when varying the signal phase $\varphi_s$, as shown in Fig.~\ref{fig7}(a). A good alignment is seen between the two curves. The fact that the gain varies with a $\pi$ periodicity of $\varphi_s$ also agrees with the theoretical prediction. When the signal and idler constructively interfere, the measured phase-sensitive gain can be as high as 22.5\,dB. The simulated phase-sensitive gain at this point is 22.0\,dB, representing a 12\,\% deviation from the measured value. As a comparison, the phase-insensitive gain at the same pump power, i.e. when the signal frequency is not at $f_c$ and the amplifier works in the non-generate mode, is only 16.0\,dB. On the other hand, when destructive interference happens for the signal and the idler, the deamplification is below -20\,dB. Similarly, the phase-sensitive gain when sweeping the phase of one of the pumps $\varphi_{p1}$ can also be extracted and compared, as plotted in Fig.~\ref{fig7}(b). In this case, the gain periodicity is $2\pi$ as predicted by Eq.~\ref{eq5}. These results demonstrate the capability of our method in accurately modeling the behavior of phase-sensitive parametric amplifiers.

\section{Conclusion and Discussion}
In conclusion, we have demonstrated the modeling of a NbN nanowire based KIPA device and compared the results with experimental data. We show that by incorporating linear kinetic inductance in the EM analysis, the linear properties of the device such as resonance frequency and external coupling strength can be accurately captured. Finding the resonance frequency at varying temperature is also made possible by including temperature-dependent London penetration depth in the EM simulation. To model the nonlinear response of the device, we implement an EM and circuit co-simulation by representing the nanowire as a circuit component. It is considered as a linear inductor in EM analysis to extract the S-parameter of the device. The S-parameter is then fed to a circuit simulator in which the nanowire is modeled as a current-dependent nonlinear inductor. The harmonic balance simulator is then exploited to extract nonlinear properties including resonance bifurcation, gain spectra, and phase-sensitive amplification. Good agreement is found between simulated and measured results.

Compared to lumped-element circuit simulations, our co-simulation approach has some unique advantages. Firstly, the device response can be accurately captured by directly exploiting the S-parameter results from the EM analysis. It is also more efficient for designing complex device structures, for which constructing an equivalent circuit model can be challenging. Furthermore, our method would be more broadly applicable to devices with distributed linear inductance end capacitance.

For future works, the nonlinear simulation can be straightforwardly extended to perform more detailed studies of the amplifiers' properties, such as gain compression, intermodulation products, and higher order harmonics. These characteristics would provide more insight for optimizing the device performance. The phase-sensitive behavior of the KIPA can be further modeled and investigated for squeeze state generations.

Our design workflow can also be applied to other kinetic inductance based devices including traveling-wave amplifiers, frequency converters, and three-wave mixing amplifiers. To model kinetic inductance traveling-wave amplifiers, the layout of the repeating structure, i.e. unit cells, can be simulated with EM analysis. The circuit co-simulation can then be performed by incorporating the nonlinear inductance and cascading the unit cells to construct the transmission line. For frequency converter devices, harmonic balance simulator can be similarly implemented to extract the mixing tones of different orders of signal and pump harmonics. These future studies can establish our approach as a widely-applicable tool for designing superconducting kinetic inductance systems.

\section*{Acknowledgment}
The experimental studies presented in this work are supported by funding from the DARPA under Grant No. HR0011-24-2-0346 and Office of Naval Research on the development of nitride-based superconductors (under Grant No. N00014-20-1-2126) . The part of the research that involves cryogenic instrumentation was supported by the US Department of Energy Co-design Center for Quantum Advantage (C2QA) under Contract No. DE-SC0012704. The authors acknowledge the Keysight EDA development team for their contribution in enabling the simulation workflow in Keysight ADS.

\bibliographystyle{IEEEtran}

\bibliography{IEEEabrv,reference}

\end{document}